\documentclass[aps,onecolumn,preprint,superscriptaddress]{revtex4-1}
\usepackage[T1]{fontenc}
\usepackage{graphicx}
\usepackage{amsmath,amssymb,cases}
\usepackage{bm}

\renewcommand{\d}{\delta}
\DeclareMathOperator{\tr}{tr}
\newcommand{\ds}{\displaystyle}
\newcommand{\ie}{{\it i.e.}}

\newcommand{\supralapl}{\mathcal{L}}
\newcommand{\suprajac}{\mathcal{J}}
\newcommand{\supraid}{I}
\newcommand{\supradeg}{\mathcal{D}}
\newcommand{\supraq}{\mathcal{Q}}

\newcommand{\suprazero}{\text{\large0}}
\newcommand{\supravec}{{\mathbf w}}
\newcommand{\lu}{L_{ij}^{(u)}}
\newcommand{\lv}{L_{ij}^{(v)}}

\newcommand{\gu}{G^{(u)}}
\newcommand{\gv}{G^{(v)}}
\newcommand{\du}{D^{(u)}}
\newcommand{\dv}{D^{(v)}}
\newcommand{\su}{\sigma^{(u)}}
\newcommand{\sv}{\sigma^{(v)}}
\newcommand{\ku}{k^{(u)}}
\newcommand{\kv}{k^{(v)}}
\newcommand{\iu}{i^{(u)}}
\newcommand{\iv}{i^{(v)}}

\begin{document}
%
%
%
\title{Pattern formation in multiplex networks}
\author{Nikos E. Kouvaris} 
\email{nikos.kouvaris@ub.edu}
\affiliation{Department of Physics, University of Barcelona, Mart\'i i Franqu$\grave{e}$s 1,E-08028, Barcelona, Spain}
\author{Shigefumi Hata}
\affiliation{Department of Mathematical Science and Advanced Technology, 
Japan Agency for Marine-Earth Science and Technology, 236-0001 Kanagawa, Japan}
\author{Albert  D\'iaz-Guilera}
\affiliation{Department of Physics, University of Barcelona, Mart\'i i Franqu$\grave{e}$s 1,E-08028, Barcelona, Spain}
%
%
%
%
\date{\today}
\begin{abstract}
The advances in understanding complex networks have generated 
increasing interest in dynamical processes occurring on them. 
Pattern formation in activator-inhibitor systems has been studied 
in networks, revealing differences from the classical continuous 
media. Here we study pattern formation in a new framework, 
namely multiplex networks. 
These are systems where activator and inhibitor species occupy 
separate nodes in different layers. Species react across layers but 
diffuse only within their own layer of distinct network topology. 
This multiplicity generates heterogeneous patterns with significant 
differences from those observed in single-layer networks. 
Remarkably, diffusion-induced instability can occur even if the 
two species have the same mobility rates; condition which can 
never destabilize single-layer networks.
The instability condition is revealed using perturbation theory 
and expressed by a combination of degrees in the different layers. 
Our theory demonstrates that the existence of such topology-driven 
instabilities is generic in multiplex networks, providing a new mechanism 
of pattern formation. 
\end{abstract}
\maketitle
%
%
%
%
%
Distributed active media support a variety of self-organized patterns, 
such as stationary and oscillatory structures, spiral waves, and 
turbulence \cite{turing,murray-book,mik-book-I}. Such media are 
often described by reaction-diffusion systems and consist of elements 
obeying an activator-inhibitor dynamics with local coupling. 
In his pioneering paper \cite{turing}, Turing showed that a uniform 
steady state can be spontaneously destabilized, leading to a 
spontaneous formation of a periodic spatial pattern, when reacting 
species diffuse with different mobilities. It was later proposed by 
Gierer and Meinhardt \cite{MEIN00} that an activator-inhibitor 
chemical reaction is a typical example achieving Turing's scenario.
Turing instability is a classical mechanism of self-organization 
far from equilibrium, and plays an important role in biological 
morphogenesis. It has been extensively studied in biological 
\cite{MEIN00,Sick2006,Maini06} and chemical \cite{Ouyang1991} 
systems, as well as real ecosystems \cite{Rietkerk2008,Liu2013}.
\par
The active elements can also be coupled in more complicated 
ways, forming complex networks \cite{Karlsson2002,Bignone2001}. 
Complex networks are ubiquitous in nature \cite{barrat-book-08}; 
two typical examples are epidemics spreading over transportation 
systems \cite{COL06} and ecological systems where distinct habitats 
communicate through dispersal connections 
\cite{Hanski1998,Urban2001,Fortuna2006,HOL08}. 
Theoretical studies of reaction-diffusion processes on complex 
networks have recently attracted much attention 
\cite{barrat-book-08,Bocaletti2006,Arenas2008,KOU12,KOU14a}. 
Othmer and Scriven \cite{oth71,oth74} developed the general 
mathematical framework to describe Turing instability in networks, 
and provided several examples of small regular lattices. Afterwards, 
Turing patterns were explored in small networks of chemical reactors 
\cite{hor04,MOO05}. More recent work in this area includes detailed 
studies of Turing bifurcation and related hysteresis phenomena in 
large complex networks \cite{Nakao2010,Wolfrum2012a}, and oscillatory 
Turing patterns in multi-species ecological networks \cite{Hata2014}.
\par
In nature, the active elements of a system can communicate 
through different types of pathways with different architecture.
Such a system with multiple types of links can be represented 
as a special type of complex network called a {\it multiplex network}
\cite{Boccaletti2014}. Recent theoretical studies have shown that the 
spectral properties of multiplex networks are significantly different from 
those of single-layer networks 
\cite{Gomez2013,SOL13,dedomenico2013,Radicchi2013,Boccaletti2014}, 
and that these differences affect the diffusion processes occurring 
on the network \cite{Gomez2013,SOL13}. Consequently, the emergent 
dynamics can exhibit new kinds of patterns. Examples include the breathing 
synchronization of cross-connected phase oscillators \cite{Louzada2013} 
and the emergence of a metacritical point in epidemic networks, where 
diffusion of awareness is able to prevent infection and control the spreading 
of a disease \cite{Granel2013}. Turing patterns have also been discussed
in the context of multiplex networks \cite{ASL14}.
\par
It has been reported that many man-made networks and real ecosystems 
are spatially fragmented in such a way that different species can migrate 
using different paths in separate layers 
\cite{Stegeman2002,Fang2009,Xuan2013,Cardillo2013a,Buono2014}. 
In studies of classical swine fever, for example, it was found that an 
individual can spread the infection by different types of contacts 
characterized by different infection rates \cite{Stegeman2002}. 
Moreover, the role of different but overlapping transportation 
networks was considered in a study exploring the diffusion 
pattern of severe acute respiratory syndrome near Beijing 
\cite{Fang2009}.
\par
This literature leads us to consider a new class of dynamical 
systems, {\it multiplex reaction networks}, where reacting species 
are transported over their own networks in distinct layers, but can 
react with each other across the inter-layer connections. This paper 
provides a general framework for multiplex reaction networks and 
constructs a theory for self-organized pattern formation in such 
networks. As a typical example, we investigate a diffusively-coupled 
activator-inhibitor system where Turing patterns can develop.

%
%
%
%
\section{Multiplex reaction networks}

\begin{figure}[htb]
\begin{center}
\includegraphics{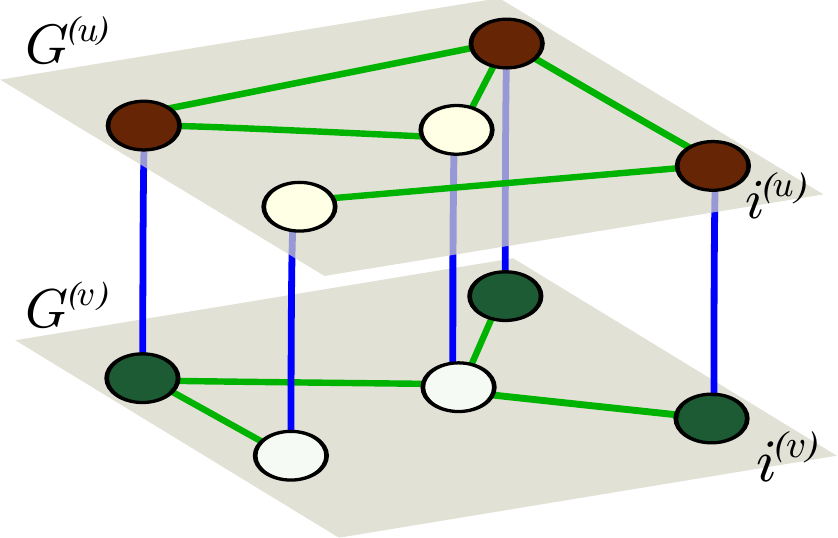}
\end{center}
\caption{{\bf Activator-inhibitor system organized in multiplex network.} 
Activator and inhibitor species occupy nodes in separate layers $\gu$ 
and $\gv$, respectively. 
They react across the layers (blue inter-layer links), 
while they migrate within their own layers (green intra-layer links).}
\label{fig1}
\end{figure}

We consider multiplex networks of activator and inhibitor populations, where
the different species occupy separate network nodes in distinct layers. Species 
react across layers according to the mechanism defined by the activator-inhibitor 
dynamics, and diffuse to other nodes in their own layer through connecting links 
(see figure~\ref{fig1}). Such a process can be described by the equations

\begin{subequations}
\begin{eqnarray}
\frac{d}{dt}u_i(t) &=& f(u_i,v_i) +  \su \ds \sum_{j=1}^{N}\! \lu u_j\,,\label{eq:rdneta}\\[2pt]
\frac{d}{dt}v_i(t) &=& g(u_i,v_i) + \sv \ds \sum_{j=1}^{N}\! \lv v_j\,,\label{eq:rdnetb}
\end{eqnarray}
\label{eq:rdnet}
\end{subequations}

\noindent where $u_i$ and $v_i$ are the densities of activator 
and inhibitor species in nodes $\iu$ and $\iv$ of layers $\gu$ and 
$\gv$, respectively.
The superscripts $(u)$ and $(v)$ refer to activator and inhibitor.
The activator nodes are labeled by indices $i=1,2,\ldots,N$ in order
of decreasing connectivity. 
The same index ordering is applied to the inhibitor layer. 
The functions $f(u_i,v_i)$ and $g(u_i,v_i)$ specify 
the activator-inhibitor dynamics. The Laplacian matrices
$L^{(u)}$ and $L^{(v)}$ describe diffusion processes 
in the two layers, and the constants $\su$ and $\sv$ are 
the corresponding mobility rates (see details in the Methods section).
\par
As a particular example we consider the Mimura-Murray 
ecological model \cite{Mimura1978} on a multiplex network 
consisting of two scale-free layers. In the absence of diffusive 
coupling, such that $\su=0$ and $\sv=0$, the multiplex system 
relaxes to a uniform state, \ie{} $(u_i, v_i)=(u_0, v_0)$ 
for all $i=1,\ldots,N$. 
The homogeneous densities are determined by $f(u_0, v_0)=g(u_0, v_0)=0$
(see Methods). Under certain conditions, which we present 
here, non-uniform patterns can evolve from an instability driven 
by the multiplex structure. 

%
%
%
%
\section{Linear stability of the uniform state}

In simplex networks, where $L^{(u)}\equiv L^{(v)}$, 
the uniform state may undergo 
a Turing instability as the ratio $\sv/\su$ increases and exceeds 
a certain threshold. The instability leads to the spontaneous emergence 
of stationary patterns consisting of nodes with high or low densities 
of activators \cite{Nakao2010}.
Such diffusion-induced instability can also take place in multiplex reaction 
networks~(\ref{eq:rdnet}). This phenomenon can be explained through a linear 
stability analysis with non-uniform perturbations.
We introduce small perturbations, $\d u_{i}$ and $\d v_{i}$, to the 
uniform steady state, as follows: $(u_{i},v_{i})=(u_0,v_0)+(\d u_{i},\d v_{i})$. 
We then substitute the perturbed state into equations \eqref{eq:rdnet} to obtain a 
set of coupled linearized differential equations. 
Finally, by means of an approximation technique
described fully in the Methods section, we obtain a characteristic equation 
for the growth rate $\lambda$ of the perturbations for each pair of nodes.
\par
The onset of the instability occurs when $\text{Re}\,\lambda=0$ 
for some pair of nodes $\iv$ and $\iu$.
The instability condition is fulfilled when these 
nodes possess a combination of degrees $\ku$ and $\kv$ 
such that, the equation

\begin{equation}
\ku = \frac{f_u g_v - f_v g_u - f_u \sv \kv } {g_v \su - \su \sv \kv}\,,
\label{eq:turing_instability}
\end{equation}

\noindent is satisfied. Here, $f_u,f_v,g_u$ and $g_v$ are partial derivatives 
at the uniform steady state.
Condition~(\ref{eq:turing_instability}) implies that a sufficiently 
large value of $\sigma^{(v)}$ brings about instability, in the same manner 
as the Turing instability. However, an alternative scenario of the instability is 
revealed by Eq.~(\ref{eq:turing_instability}). This can happen
by increasing $\kv$, even if the mobilities are equal ($\su = \sv$).
This instability occurs in a strikingly different regime from classical 
diffusion-induced instabilities.

\begin{figure*}[htb]
\begin{center}
\includegraphics{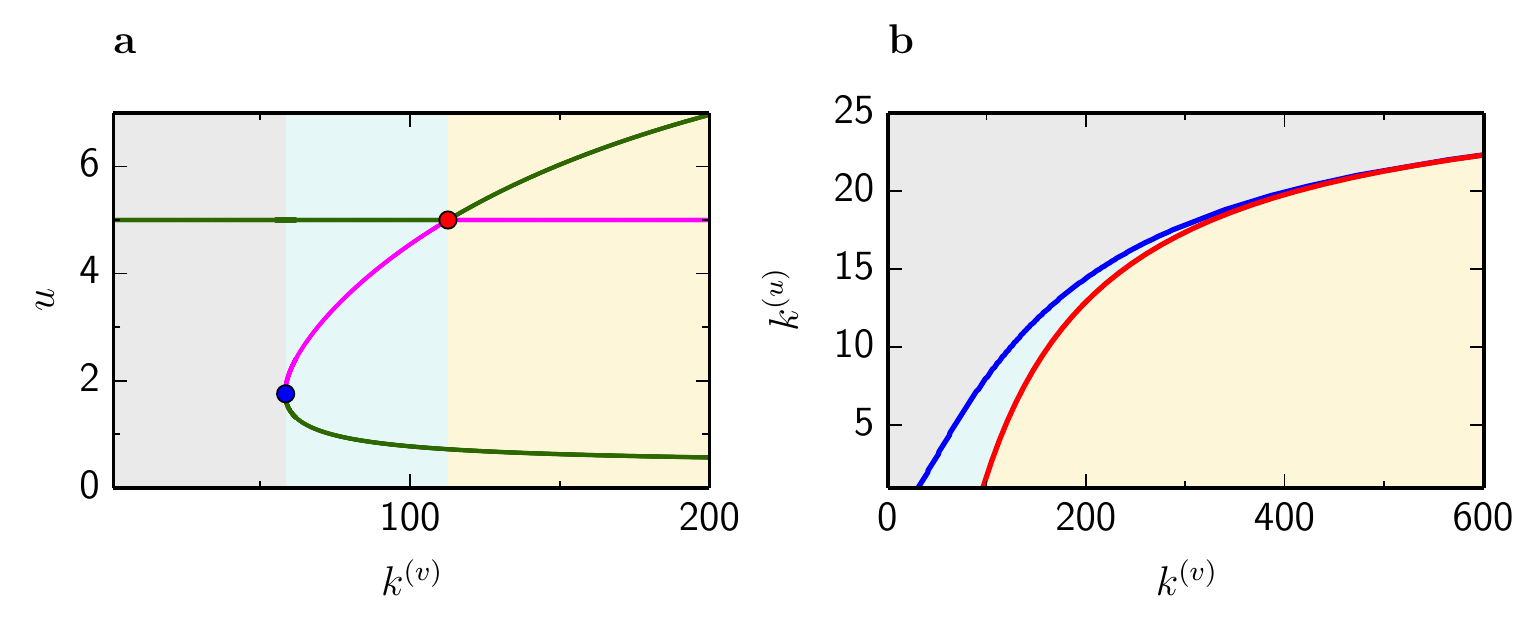}
\end{center}
\caption{{\bf Bifurcation diagram.} 
{\bf a,} Stationary solutions of system~\eqref{eq:rdnet} for $\ku=4$. 
Green curves indicate stable solutions while magenta curves 
correspond to unstable solutions of the linearized system. 
Red point indicates the transcritical bifurcation where the uniform steady 
state $(u_0,v_0)=(5,10)$ becomes unstable. Blue point corresponds 
to a saddle-node bifurcation of a solution $(u,v)$ which originates from 
the transcritical bifurcation. 
{\bf b,} Transcritical bifurcation (red curve) given by Eq. \eqref{eq:turing_instability}, 
is shown together with the continuation of the saddle-node bifurcation (blue curve) in the 
plane $\kv$-$\ku$.}
\label{fig2}
\end{figure*}

Figure \ref{fig2}a shows the linear stability of system \eqref{eq:rdnet} 
for varying $\kv$, holding $\ku$ fixed. We clearly see that the 
uniform steady state is always a solution of the multiplex system. 
It is linearly stable (green line) for small values of $\kv$. But at 
some critical value of $\kv$ which satisfies equation \eqref{eq:turing_instability},  
the system undergoes a transcritical bifurcation (red point) and becomes unstable 
(magenta line). Two new branches of solutions arise from the transcritical bifurcation.
The unstable branch (magenta line) undergoes a second bifurcation (blue point), 
this time a saddle-node, giving rise to a new branch of stable 
solutions (green line) different from the uniform steady state.
Figure \ref{fig2}b shows the transcritical (red line) and the saddle-node 
bifurcation (blue line) in the $\kv$-$\ku$ plane. The curve of the transcritical bifurcation  
is given by equation \eqref{eq:turing_instability}, while the curve of the saddle-node 
bifurcation has been derived by numerical continuation.
One can see from Eq.~(\ref{eq:turing_instability}) that by increasing $\kv$, 
the boundary curve (red line) asymptotically approaches $\ku=f_u/\su$. 
This indicates that the instability can be observed with sufficiently large $\kv$, 
if the mean degree of the activator layer is less than this value. 
This fact reveals an important difference from the classical Turing instability, 
which always takes place by increasing $\sv$ irrespective of $\su$ \cite{Nakao2010}.

\par
The diffusion-induced instability occurs on the transcritical bifurcation. However, 
non-uniform patterns can also develop after the saddle-node 
bifurcation. In other words, we find that multiplex systems exhibit multistability in the area
between these two bifurcations (cyan), where a branch of stable solutions
coexists with the uniform steady state.

%
%
%
%
\section{Pattern formation arising from the instability}
Suppose that the multiplex system starts almost in 
the uniform steady state with small perturbations.
Equation \eqref{eq:turing_instability} allows us to 
identify pairs of nodes ($\iv$, $\iu$) where the
small perturbations will be amplified, 
so that these nodes leave the uniform state, 
triggering the formation of a non-uniform stationary 
pattern. Such a pattern cannot develop from pairs of 
nodes possessing degrees in the grey area of Figure \ref{fig2}b,
where only the uniform state exists. 
However, pairs of nodes with degrees in the yellow area, 
beyond the transcritical bifurcation, are unstable. 
Under small perturbations they can leave the uniform state,
yielding the formation of a stationary non-uniform pattern. 
The cyan area between the two bifurcations indicates that 
the system exhibits multistability, where the uniform steady 
state coexists with a branch of solutions corresponding to 
non-uniform patterns. 

\begin{figure*}[htb]
\begin{center}
\includegraphics{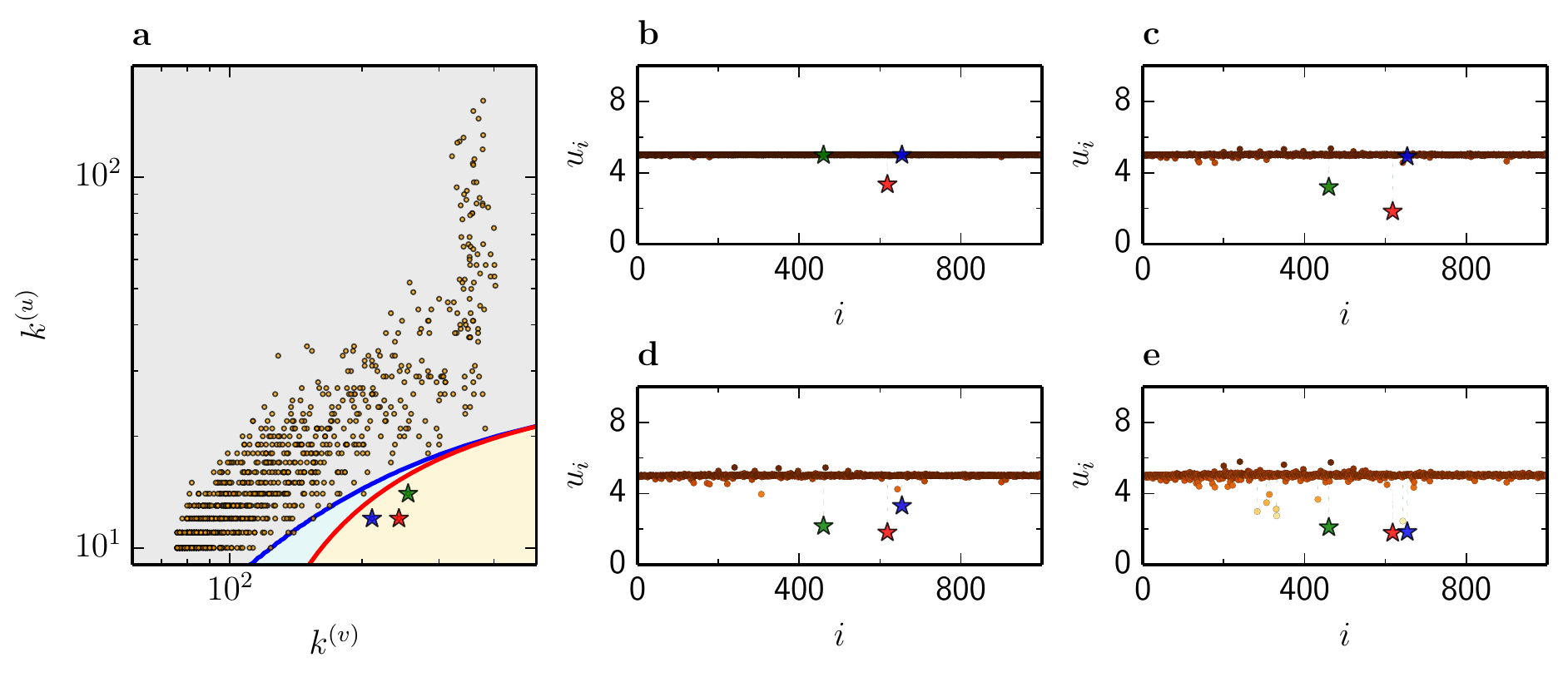}
\end{center}
\caption{{\bf Multiplex diffusion-induced instability.} 
{\bf a,} Degree combination for pairs of nodes $\iv$ and $\iu$ is shown 
in plane $\kv$-$\ku$ together with the curves of saddle-node (blue) 
and transcritical (red) bifurcations. 
Snapshots of the activator pattern for
$t=50$ ({\bf b}), 
$t=63$ ({\bf c}),
$t=70$ ({\bf d})
and the fully developed pattern for 
$t=500$ ({\bf e})
are shown for the Mimura-Murray model with $\sv=\su=0.12$ on a multiplex 
network with scale-free layers of $N=1,000$ nodes and mean 
degrees $\langle\kv\rangle=152$ and $\langle\ku\rangle=20$
(see also Supplementary Movie S1).
Nodes are ordered according to decreasing degrees $\ku$.}
\label{fig3}
\end{figure*}

We verify this scenario for a multiplex network where both 
layers, $\gv$ and $\gu$, are scale-free. Figure \ref{fig3}a
displays the actual degree combination ($\kv$, $\ku$) for 
each pair of nodes $\iv$, $\iu$ (orange points) of this network
in the $\kv$-$\ku$ plane, together with the bifurcation curves. 
Three pairs of nodes, the critical ones, denoted by 
stars, have degrees exceeding the instability threshold.  
Thus, a non-uniform pattern starts to grow from these 
nodes. The critical node denoted by the red star is the
first to spontaneously leave the uniform state, 
as shown in figure \ref{fig3}b.
Next, figures \ref{fig3}c,d show that the critical 
nodes denoted by the green and blue stars rapidly 
differentiate from the uniform state. 
Finally, triggered by these growing perturbations, all nodes leave 
the steady state to establish a non-uniform pattern
(\figurename~\ref{fig3}e).

\begin{figure}[htb]
\begin{center}
\includegraphics{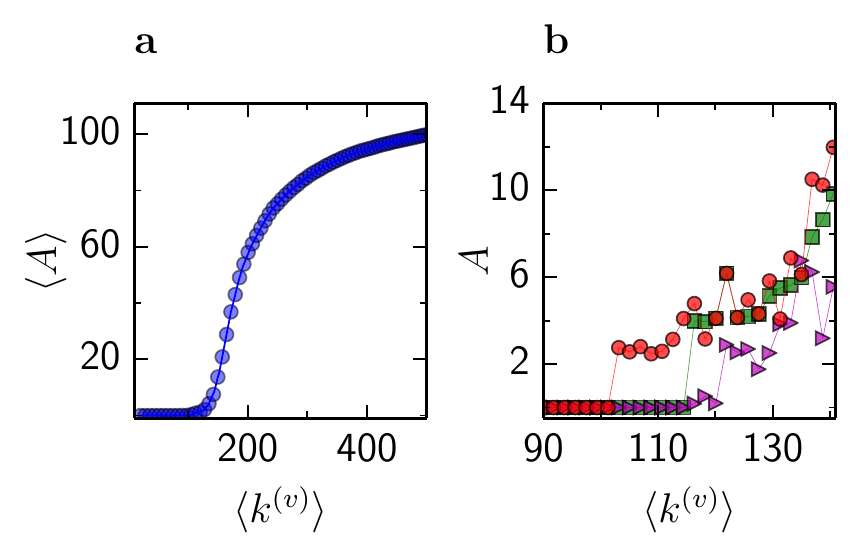}
\end{center}
\caption{{\bf Amplitude of non-uniform patterns.} 
{\bf a,}  Average amplitude of non-uniform pattern is shown 
as a function of $\langle \kv \rangle$ for 
$\langle \ku \rangle=20$ and $\su=\sv=0.12$. 
Average is taken over ten numerical simulations 
for different implementation of $\gv$ with the same 
mean degree $\langle \kv \rangle$.
{\bf b,} Amplitude in the vicinity of transition for 
three numerical simulations where different perturbations 
were applied to the same sequence of networks $\gv$.}
\label{fig4}
\end{figure}

Multistability corresponding to the cyan are of  \figurename~\ref{fig2}b
has been studied via numerical simulations.
Figure \ref{fig4}a shows the amplitude $A$ of the observed patterns 
(see Methods section), averaged over different simulations. 
Each point of the diagram is the average of ten different implementations 
of $\gv$ with the same mean degree $\langle \kv \rangle$; $\gu$ is fixed.
We clearly see that the amplitude is zero; \ie{}, the uniform state is the only 
stable attractor of the system, for $\langle\kv\rangle$ smaller than a critical 
threshold $\langle\kv\rangle_{\textrm c}$. 
However, a more detailed look in the vicinity of this transition 
reveals that a number of different stationary patterns could be 
identified for the same parameter values.
As an example, figure~\ref{fig4}b shows the amplitudes 
in three simulations where different perturbations have 
been applied to the same sequence of multiplex networks. 
Starting from the uniform state with small perturbations, 
the instability occurs at some critical threshold,  
resulting in a small abrupt increase of amplitude.
Different perturbations result in different values for the 
instability threshold.

\begin{figure*}[htb]
\begin{center}
\includegraphics{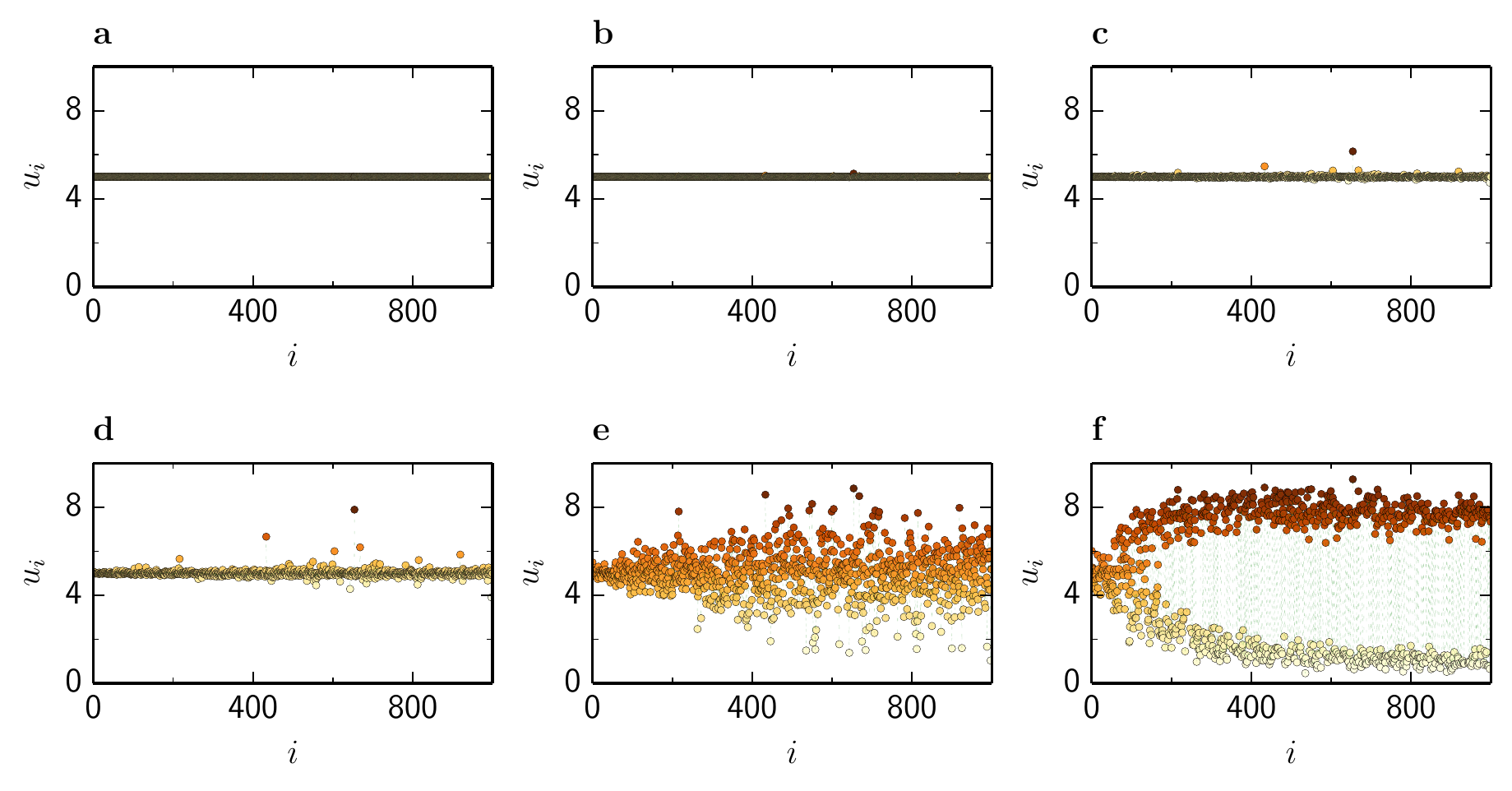}
\end{center}
\caption{{\bf Development of non-uniform pattern.} 
The Mimura-Murray model with mobilities $\sv=\su=0.12$
 on a multiplex network with scale-free layers of $N=1,000$ 
 nodes, and mean degrees $\langle\kv\rangle=500$ and 
 $\langle\ku\rangle=20$.
Small perturbations are added to the uniform steady state and
nodes that satisfy condition \eqref{eq:turing_instability} 
loose their stability and leave the uniform state.
Snapshots of the activator pattern at time 
$t=11.5$ ({\bf a}),
$t=13.5$ ({\bf b}), 
$t=15$ ({\bf c}),
$t=16$ ({\bf d}), 
$t=18$ ({\bf e})
and
$t=500$ ({\bf f})
are shown (see also Supplementary Movie S2).
Nodes are ordered according to decreasing degrees $\ku$.}
\label{fig5}
\end{figure*}

\begin{figure*}[htb]
\begin{center}
\includegraphics{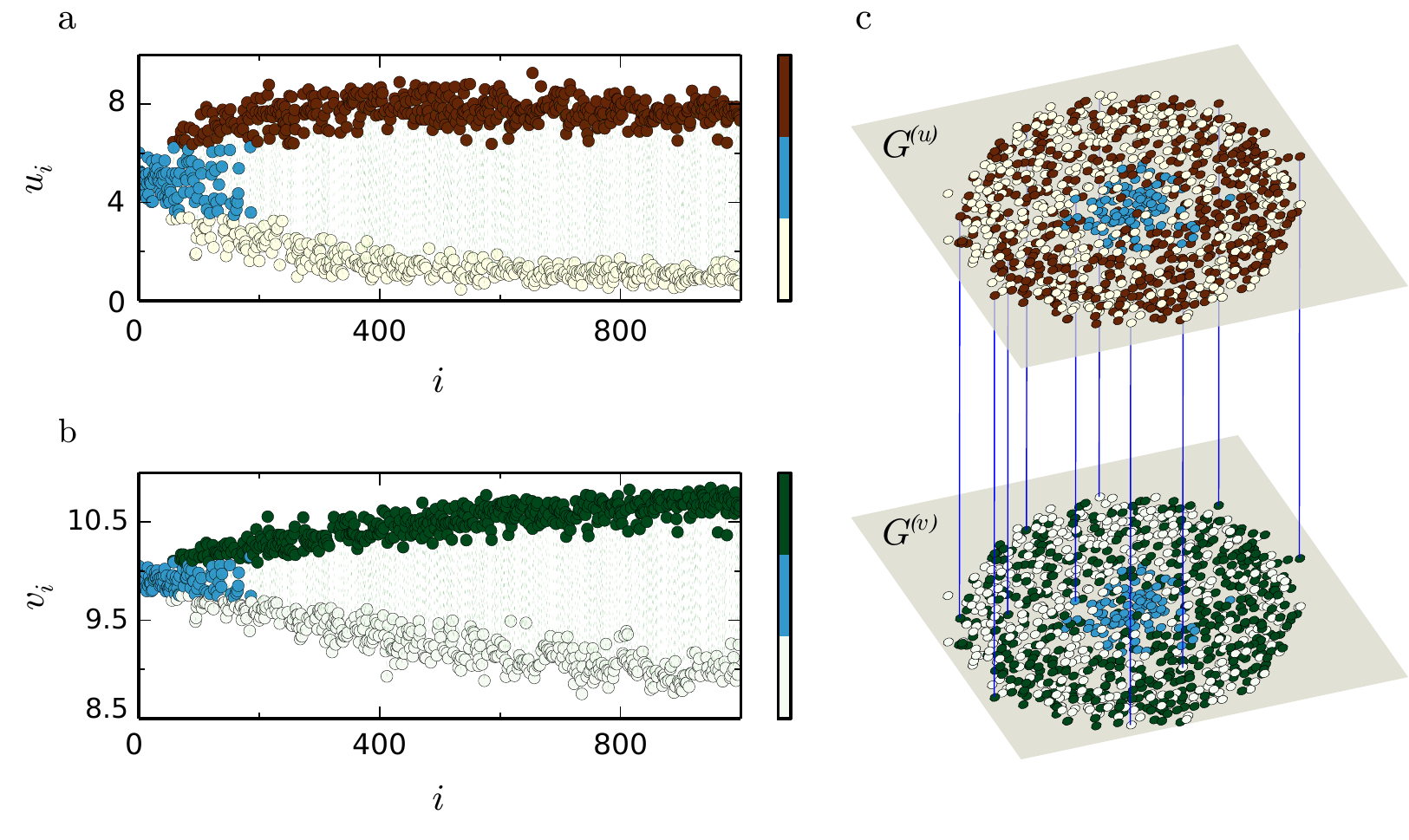}
\end{center}
\caption{{\bf Actual multiplex pattern.} 
The Mimura-Murray model with mobilities $\sv=\su=0.12$ 
on a multiplex network with scale-free layers of $N=1,000$ 
nodes, and mean degrees $\langle\kv\rangle=500$ and 
$\langle\ku\rangle=20$. 
Non-uniform stationary pattern is shown in the activator layer $\gu$ ({\bf a}) 
as well as in the inhibitor layer $\gv$ ({\bf b}). Nodes in activator layer are 
ordered according to decreasing degree; nodes in the inhibitor layer are 
ordered correspondingly. 
{\bf c,} Same pattern is shown in the actual multiplex network. 
Nodes in $\gu$ are plotted using a spring algorithm, so that,
those having high degrees are placed in the center and 
those with small degrees in the periphery. Nodes in $\gv$
follow the same indexing. 
For convenience, intra-layer links are not shown, while 
from inter-layer links only few are chosen to be shown.}
\label{fig6}
\end{figure*}

Obviously, different $\langle\kv\rangle$ values lead to patterns of different 
amplitudes. Figure \ref{fig3}e shows a pattern for $\langle\kv\rangle=152$, 
close to the transition. However, patterns 
where more nodes leave the uniform state can also develop far from the
transition. Figures \ref{fig5}a-f show the evolution of small perturbations 
in the uniform state and the formation of a non-uniform pattern in a 
multiplex network with scale-free layers of $N=1,000$ nodes, 
and mean degrees $\langle\kv\rangle=500$ and $\langle\ku\rangle=20$. 
Under the influence of small perturbations, some critical 
nodes differentiate rapidly from the uniform steady state.
Afterwards, nonlinear effects (which are not 
described by our theory) drive the multiplex system to 
self-organize into a stationary pattern with two separate 
group of nodes. 
The separation between nodes of low and high activator densities 
is more pronounced in nodes with small degrees 
$\ku$, while nodes with large $\ku$ tend to sustain their initial state. 
Figures \ref{fig6}a,b show this pattern in the activator and inhibitor
layers respectively, whereas figure \ref{fig6}c shows the actual multiplex
pattern.

%
%
%
%
\section{Discussion}

We have proposed a new class of dynamical systems, multiplex reaction
networks, where each reacting species occupies its own network layer 
and reacts with the other species using cross-layer contacts. 
As a demonstration of this new reaction scheme, we investigate pattern 
formation induced by diffusive transport in a multiplex network with two reacting 
species. Our theory, based on linear stability analysis with perturbations around the
uniform steady state, correctly predicts the instability threshold observed in numerical 
simulations of the multiplex network.
\par
If the different layers have the same architecture, \ie{} $L^{(u)}=L^{(v)}$,
then this multiplex diffusion-induced instability reduces to the well-known Turing instability
which may occur when the inhibitor diffuses much faster than the activator.
Our theory~(\ref{eq:turing_instability}) predicts that the analogous instability 
can also appear in multiplex reaction networks by increasing the inhibitor diffusion rate. 
However, a significantly different kind of instability can occur in multiplex 
reaction networks, even if the two species have the same mobilities ($\su=\sv$). 
This new instability is related to the degree combination ($\kv$, $\ku$) of a pair of nodes.
The basic condition for any given pair of nodes $\iv$, $\iu$ to undergo instability 
is that their degrees $\kv$ and $\ku$ satisfy Eq.~(\ref{eq:turing_instability}).
Indeed, the instability always takes place for any $\ku$ which is less or 
equal to the value calculated from Eq.~(\ref{eq:turing_instability}), for a given 
large $\kv$. 
\par
Similar to simplex networks, multiplex systems exhibit multistability. 
The onset of pattern formation can occur even before the instability
described by Eq.~(\ref{eq:turing_instability}).
The minimal condition for developing non-uniform
patterns is that in a pair of nodes $\iv$, $\iu$ the degree $\ku$ is less 
than or equal to the value on the saddle-node 
bifurcation curve that corresponds to $\kv$. 
In the multistability regime, different stationary patterns can coexist 
with the uniform steady state for the same parameter values. 
\par
Although the observed properties of the stationary patterns are similar 
to those found in simplex networks \cite{Nakao2010,Wolfrum2012a}, 
the cause of destabilization of the uniform state is different.
This cause is only characteristic of multiplex networks and lies in the relationship 
between $\kv$ and $\ku$ for a pair of nodes. 
Therefore, the purposeful design of nonequilibrium patterns 
should be possible by tuning the architecture of the multiplex structure. 
Recently, new algorithms for building multiplex networks with 
positive or negative degree correlations across the layers have been 
proposed \cite{Lee2012,Nicosia2013b,Kim2013}. Using these algorithms, 
we can design multiplex networks where the onset of instability is 
controlled by tuning the degrees $\kv$ and $\ku$,
and the source of instability can be located at any desired pair of nodes $\iv$, $\iu$.
\par
Multiplex networks can be used to represent different types of interaction 
\cite{Stegeman2002, Granel2013,Buono2014} or different transportation
lines \cite{Fang2009,Kaluza2010,Cardillo2013a} between discrete nodes. 
In ecological multiplex networks, for example, pairs of nodes might represent separate
habitat patches which communicate through dispersal connections. 
However, prey and predators may use different connections (such as forest 
paths, rivers and tributaries or various transportation systems) to 
move among the fragmented habitats. Often, predators have more 
choices to move; in our representation their layer is 
more densely connected than the prey's layer. 
This is exactly the sort of situation that favors the formation of instability 
and the subsequent establishment of non-uniform patterns. Considering 
that self-organized patterns can be found in real ecosystems \cite{Rietkerk2008,Liu2013} 
it is possible that such patterns can also be observed in natural ecological 
systems for which the multiplex structure is innate.

%
%
%
%
\section{Methods}

\paragraph*{\bf Layer architecture.} 
In the numerical simulations, each layer is a scale-free 
network constructed by the preferential attachment algorithm 
\cite{Albert2002}. The network structure is determined by a 
symmetric adjacency matrix $A$, whose elements $A_{ij}$ 
are $1$ if there is a link connecting nodes $i$ and $j$, and 
$0$ otherwise. The degree, \ie{} number of links, of node $i$ 
is defined as $k_i = \sum_{j=1}^N A_{ij}$. The network 
Laplacian matrix $L$ is given by the expression 
$L_{ij} = A_{ij} - k_i \d_{ij}$. 
\par
The activator's network $\gu$ was constructed with mean 
degree $\langle \ku \rangle=20$. The same network was 
used throughout all numerical simulations. Each simulation 
uses a different realization of the inhibitor's network $\gv$, 
whose mean degree $\langle \kv \rangle$ is varied between 
simulations. The superscripts $(u)$ and $(v)$ refer to activator 
and inhibitor. For convenience, the indices $\iu$ of nodes in 
the layer $\gu$ are assigned in order of decreasing degrees 
$\ku_i$: that is, $\ku_1\ge \ku_2\ge \cdots \ge \ku_N$. The 
nodes $\iv$ in the layer $\gv$ follow the ordering of their 
counterpart in $\gu$, so for example node $1$ in the inhibitor 
network (the most highly connected node) always corresponds 
to node $1$ in the activator network, but the latter may or may 
not be highly connected.

\paragraph*{\bf Multiplex networks.} 
The multiplex networks used in our numerical simulations 
consist of two separate layers and two different types of 
links, {\it intra-layer} and {\it inter-layer} links. {\it Intra-layer} 
links are described by the adjacency matrices and limit the 
diffusional mobility of the species. {\it Inter-layer} links connect 
every node $i^{(u)}$ of layer $\gu$ to its counterpart $i^{(v)}$ 
in layer $\gv$. They represent the reaction dynamics defined 
in the functions $f(u_i,v_i)$ and $g(u_i,v_i)$.

\paragraph*{\bf Activator-inhibitor dynamics.}
We choose the Mimura-Murray model \cite{Mimura1978} 
as an example of an activator-inhibitor system. In this 
model the dynamics are given by the functions 
$f(u,v) = \left[\left(a+bu-u^2\right)/c-v\right]u$ 
and $g(u,v) = (u - dv -1) v$, where $u,v$ correspond 
to the densities of activator and inhibitor respectively. 
The chosen parameters are $a=35,\ b=16,\ c=9,\ d=0.4$,
yielding the linearly stable fixed point $(u_0, v_0)=(5,10)$. 
This requires the networks to satisfy $\tr(J_{(u_0,v_0)})<0$ 
and $\det(J_{(u_0,v_0)})>0$, where $J$ is the Jacobian matrix 
$ J_{(u,v)} = 
\left(
\begin{array}{cc}
f_u   & f_v  \\
g_u  & g_v  \\
\end{array}
\right)
$,
and 
$f_{u} = \partial f / \partial u$, 
$f_{v} = \partial f / \partial v$,
$g_{u} = \partial g / \partial u$ 
and
$g_{v} = \partial g / \partial v$
are partial derivatives.

\paragraph*{\bf Linear stability analysis.}
The linear stability analysis is performed using a perturbation method.
We introduce small perturbations $(\d u_{i},\d v_{i})$ to the uniform 
steady state $(u_0,v_0)$, as $(u_{i},v_{i})=(u_0,v_0)+(\d u_{i},\d v_{i})$.
Substituting into equations \eqref{eq:rdnet}, we obtain the linearized 
differential equations 
$d \d u_i/dt = f_u\d u_i + f_v\d v_i +  \su\sum_{j=1}^{N}\! \lu\d u_j\,$ 
and 
$d \d v_i/dt = g_u\d u_i + g_v\d v_i + \sv\sum_{j=1}^{N}\! \lv\d v_j\,$.
Alternatively, the linearized differential equations can be written as
$d \supravec / dt = (\suprajac + \supralapl) \supravec$, 
where
$\supravec=(\d u_1,\cdots, \d u_N,\d v_1,\cdots, \d v_N)^T$ 
is the perturbation vector, 
$ \suprajac_{(u,v)} = 
\left(
\begin{array}{cc}
f_u \supraid  & f_v \supraid \\
g_u \supraid   & g_v \supraid \\
\end{array}
\right) 
$ and $ 
\supralapl = 
\left(
\begin{array}{cc}
\su L^{(u)}   & \suprazero \\
\suprazero  & \sv L^{(v)} \\
\end{array}
\right)$;
$\supraid$ is the $N\times N$ identity matrix.
For the linear stability analysis, the perturbation vector $\supravec$ 
should be expanded over the set of eigenvectors of the matrix 
$\supraq = \suprajac +  \supralapl$. It is, however, difficult to calculate 
them for different network topologies, \ie{} different Laplacian matrices 
$L^{(u)}$ and $L^{(v)}$.
Here we propose an approximation technique to analyze the 
linear stability of the system. Matrix $\supralapl$ is split into 
$\supralapl =  \supraq_0-\supradeg$, 
where
$ \supraq_0 = 
\left(
\begin{array}{cc}
\su A^{(u)}    & \suprazero \\
\suprazero  & \sv A^{(v)}  \\
\end{array}
\right)
$
and
$\supradeg = 
\left(
\begin{array}{cc}
\su\du   & \suprazero \\
\suprazero   & \sv \dv  \\
\end{array}
\right) $.
The matrices $A^{(u)}$ and $A^{(v)}$ are the adjacency
matrices of layers $\gu$ and $\gv$, respectively. The 
matrices $\du$ and $\dv$ are the corresponding degree 
matrices, which have the nodes degrees in the main 
diagonal and are zero elsewhere. 
Then, matrix $\supraq$ can be rewritten as 
$\supraq = \supraq_0 + \supraq_1$, where
$\supraq_1 = 
\left(
\begin{array}{cc}
f_u \supraid - \su\du  & f_v \supraid \\
g_u \supraid   & g_v \supraid - \sv\dv \\
\end{array}
\right) $.
Examining matrices 
$\supraq_0$ and $\supraq_1$, 
the first has elements with values of order 
$\mathcal{O}(\su)$ or $\mathcal{O}(\sv)$, 
while the second has elements with values of order 
$\mathcal{O}(\su\langle \ku \rangle )$ 
or 
$\mathcal{O}(\sv\langle \kv \rangle )$.
If both layers are dense enough that 
$\langle \ku \rangle \gg 1$ 
and
$\langle \kv \rangle \gg 1$,
we can clearly see that the elements of matrix $\supraq_1$ 
have larger values than those of matrix $\supraq_0$, 
so that $\supraq_0$ can be neglected. This approximation yields 
the approximate linearized equation 
$d \supravec / dt = \supraq_1 \supravec$.
The characteristic equation for the eigenvalues 
$\lambda$ is then given by
\begin{equation}
\det\left(
\begin{array}{cc}
f_u  - \su \ku - \lambda & f_v  \\
g_u   & g_v - \sv \kv - \lambda \\
\end{array}
\right) 
= 0\,,\nonumber
\label{eq:char_eq}
\end{equation}

\noindent and is the same for each pair of nodes $\iv$, $\iu$.
\par
This approximation neglects entirely the matrix $\supraq_0$, 
which is associated with the precise architectures of the layers. 
Instead, each node is characterized only by its degree.
This is quite similar to the powerful mean-field methods 
used for analyzing Turing patterns in single-layer networks 
\cite{Nakao2010,Wolfrum2012a}, and is always valid 
for multiplex networks consisting of layers with small diameters.

\paragraph*{\bf Amplitude of non-uniform patterns.} 
The amplitude of a non-uniform pattern is quantified as
$A=\left[\sum_{i=1}^{N}\, \left\{ \left( u_i -  u_0  \right) ^ 2 + \left(v_i - v_0 \right) ^ 2 \right\}\right]^{1/2} \,$.

%
%
%
%
\renewcommand*{\bibfont}{\footnotesize}

%
%
%
\paragraph*{\bf Acknowledgments.}
This study was supported by the EU/FP7-2012-STREP-318132 in the framework project LASAGNE. 
Partial support from Spanish DGICYT Grant No. FIS2012-38266-C02-02, 
Cross-ministerial Strategic Innovation Promotion Program 
and JSPS KAKENHI (Grant No. 26880033) in Japan 
is also acknowledged.

%
%
%
\paragraph*{\bf Author Contributions}
N.E.K. and A.D.-G. conceived and designed the study. 
N.E.K. performed numerical simulations.
All authors carried out the analysis and wrote the article.

\end{document}